# Plasmonic Metamaterial Perovskite Solar Cells: Fundamental Tradeoffs, Limitations, and Opportunities

*Kwangjin Kim*[1,2] and *Seungwoo Lee*[2*]

[1]SKKU Advanced Institute of Nanotechnology, Sungkyunkwan University (SKKU), Suwon 16419, Republic of Korea

[2]KU-KIST Graduate School of Converging Science and Technology, Korea University, Seoul 02841, Republic of Korea

*Email: seungwoo@korea.ac.kr

Abstract: Whether dispersal of plasmonic nanoparticles (NPs) within a perovskite active layer can increase the efficiency of solar cells is a long-standing question. It is well known that inclusion of metallic NPs in an active layer can boost the surrounding near-field intensity around them owing to the dipolar localized surface plasmon resonance (LSPR, also called antenna effect), which can increase light absorption by solar cells. However, the use of plasmonic NPs in perovskite solar cells has been barely reported, and it is not known whether inserting plasmonic NPs into a perovskite active layer produces any performance advantage compared with a pure perovskite counterpart. We explore the fundamental and practical limits of "plasmonic metamaterial" perovskite solar cells by applying effective medium theory and a detailed balance analysis. Our results indicate that an increase in effective refractive index of perovskite through dispersed plasmonic NPs can in principle enhance the performance of solar cells.

TOC:

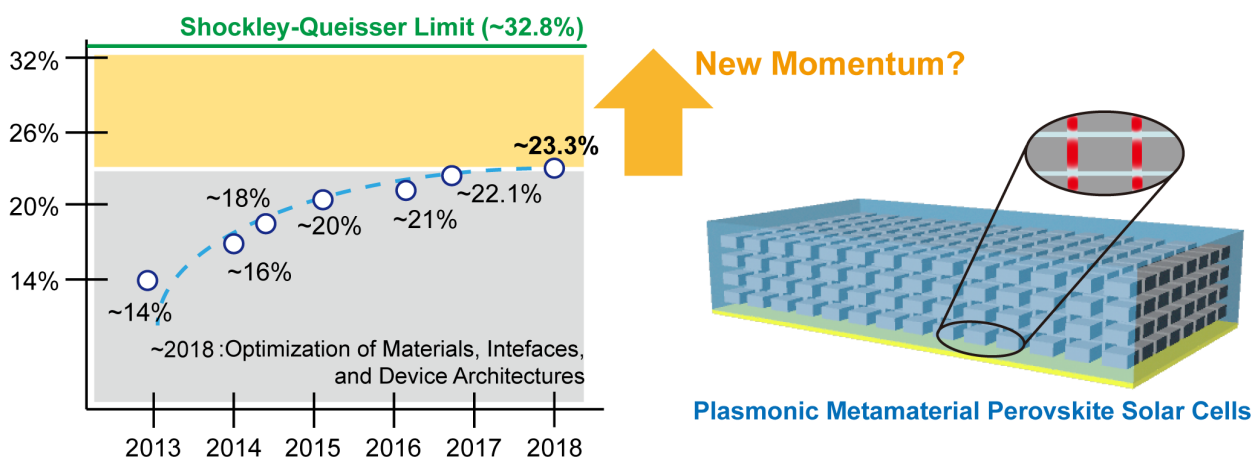

Introduction: Increasing the refractive index ($n$) has long been a goal of photon management in solar cells. The Yablonovitch limit ($4n^2$ limit) is an example of a target maximum for ray-optic light trapping;[1] S. Fan and colleagues successfully probed that the $4n^2$ limit also applies to the nanophotonic regime.[2] The recently raised claim regarding the importance of recycling radiative luminescence (also known as photon-recycling) in boosting open-circuit voltage ($V_{oc}$) has further emphasized the significance of of $n$,[3] because a higher $n$ can confine more photon gas within a medium by a factor of $n^2$.[1,3] Even if an unnaturally low $n$ (e.g., index-near-zero (INZ)) is found to afford another degree of freedom for light trapping in solar cells,[4] increasing the value of $n$ is considered key to efficient photon management in solar cells and thus enhancement of power conversion efficiency (PCE (%)).[1,2]

Meanwhile, lead halide perovskite has opened promising new opportunities for solar cell research. In particular, perovskite solar cell efficiency has experienced considerable progress over the past few years, from an initially demonstrated PCE of 12 % to the recently achieved value of 23.3 %.[5-18] Much of this advancement can be traced to optimization of lead halide materials, interfacial property, and device architecture.[5-18] However, the surging trend in PCE brought about by these engineering approaches has declined since reaching approximately 20 % (see TOC figure),[5,19] which is still far below the fundamental PCE limit (also called Shockley-Queisser limit (S-Q limit)) of lead halide perovskite solar cells as defined by detailed balance analysis.[20,21] New momentum in perovskite solar cell research is needed to approach the S-Q limit.

At the same time, photon management in perovskite solar cells via enhanced light-matter interaction has attracted comparatively less interest, probably due to their already excellent light absorption and strong photon-recycling ability.[19,22,23] However, efficient absorption of light beyond the perovskite bandgap has yet to be demonstrated. Also, a relatively low $n$ for perovskite (~ 2.5, for example, for lead halide) compared with those of gallium arsenide and silicon (3.0 ~ 3.5) would be detrimental in terms of maximizing photon-recycling. These challenges raise the question of whether an increase in effective $n$ ($n_{eff}$) of perovskite can indeed boost the PCE of solar cells.

To address this question, we studied the effect of increasing $n_{eff}$ on photon management in perovskite and the resulting solar cell performance. Deep-subwavelength-scale plasmonic nanoparticles (NPs) dispersed within perovskite can act as electrical meta-atoms, broadly enhancing $n_{eff}$ while simultaneously maintaining high transparency at a quasi-static (off-resonance) regime.[24-26] This optical effective media provides a good model system to increase the $n_{eff}$ of perovskite at a broadband solar spectrum with minimal parasitic absorption. To this end, we designed perovskite-based optical effective media using the retrieval of effective parameter (i.e., the *s*-parameter extraction method).[22-24] A detailed balance analysis[20,21,27,28] was then carried out to systematically verify the influence of enhanced $n_{eff}$ on solar cell performance. Our work provides insight into enhanced $n_{eff}$-enabled photon management of perovskite solar cells, including fundamental tradeoffs, limitations, and opportunities.

**Design of plasmonic metamaterial perovskite**: Recently, a detailed balance analysis was successfully adapted to perovskite solar cells by explaining how efficiency is influenced by non-radiative recombination (Shockley-Read-Hall (SRH) non-radiative recombination rate; $A_{SRH}=1/\tau$, where $\tau$ is non-radiative lifetime ($s^{-1}$)), Auger limit ($C_{Aug}$ ($s^{-1}$)), light absorptivity (*a*), external luminescence efficiency ($\eta_{ext}$ (%)) and parasitic absorption of luminescence ($P_{par}$).[20,21] For this detailed balance analysis on perovskite solar cells, we designed a plasmonic metamaterial perovskite effective medium (PMPEM) as follows.

The thickness of a perovskite active layer in solar cells is typically comparable to the wavelength of the solar spectrum (< 1500 nm); consequently, *a* should be interpreted under the nanophotonic regime rather than its ray-optic counterpart. However, a previously reported detailed balance analysis on perovskite solar cells thinner than 1500 nm employed ray-optical light trapping for both flat and randomly textured solar cells.[20,21] Along with these standards, we also used a ray-optic approximation for $a$. Because the main goal of our study is to elucidate the role of increasing $n_{eff}$ in solar cell performance, we hereafter include the $a$ only for a randomly textured perovskite solar cell with maximized light trapping ($4n^2$ limit):[1,20,21]

$$a(E) = \frac{4n^2\alpha d}{4n^2\alpha d + \sin^2\theta_m} \quad (1)$$

where $d$, $\theta_m$, and $\alpha$ are thickness, maximum angle of emission, and absorption coefficient, respectively. As we focused on randomly textured perovskite solar cells, $\theta_m$ is assumed to be 90°. Generally, this approximation of $a$ is effective when $\alpha d$ is much smaller than 1. The $\alpha$ of perovskite (i.e., lead halide) used in this study (see **Figure 1**a) was from previously reported experimental values;[29,30] direct and indirect bandgaps of lead halide (i.e., lead iodide) resulting from the Rashba effect[31,32] were reflected in $\alpha$. At wavelengths below the bandgap (~ 760 nm wavelength (1.62 eV)), lead halide exhibits strong light absorption (i.e., $\alpha d$ ~ 1.0 for 1000 nm thickness at 450 nm wavelength), and the light trapping effect via increasing $n_{eff}$ cannot be as effective in this regime. Enhancement of $n_{eff}$ at wavelengths larger than the bandgap should therefore be a main goal in designing a PMPEM.

As shown in **Figures S1**a, increasing the $n_{eff}$ of active layers (assumed to have $\alpha$ of lead iodide with a constant $n$ of 2.5) selectively at wavelengths larger than 760 nm (i.e., from 2.5 to 4) can considerably enhance $a$ and result in an increased short circuit current ($J_{sc}$) as follows:

$$J_{SC} = q\int_0^\infty a(E)\Phi_S(E)dE \quad (2)$$

where $q$ and $\Phi_S(E)$ are the elementary charge and solar spectrum, respectively. By contrast, increasing $n_{eff}$ selectively at wavelengths smaller than 760 nm (i.e., also from 2.5 to 4) slightly enhance $a$ and $J_{sc}$, due to the already matured light absorption of perovskite (see **Figures S1**a-b). Note that the light trapping effect via increasing $n_{eff}$ was not as well estimated in a sub-1500 nm-thick perovskite layer, because our approximation depended on ray-optics. The nanophotonic detailed balance analysis on perovskite solar cell will be reported separately, as it is beyond the scope of this study.

Another key to success in the design of a PMPEM is minimizing $P_{par}$ of luminescence (at wavelengths of 680 nm ~ 880 nm), as follows. It is well known that $\eta_{ext}$ is a determinant for $V_{oc}$ according to the following equation derived by Ross in 1967[28]:

$$qV_{oc} = qV_{oc,max} - kT|\ln\eta_{ext}| \quad (3)$$

where $V_{oc,max}$ is the maximum limit of $V_{oc}$, defined by the quasi-Fermi level. Smaller $\eta_{ext}$ results in reduced $V_{oc}$ and thus lower cell efficiency. According to the following equation, $\eta_{ext}$ is reduced by $P_{par}$:[21]

$$\eta_{ext} = \frac{B_{int}P_{esc}i^2}{A_{SRH}i + B_{int}(P_{esc}+P_{par})i^2 + C_{Auger}i^3} \quad (4)$$

where $P_{esc}$, $i$, and $B_{int}$ are escape probability, available carrier density, and internal radiative coefficient, respectively. Plasmonic NPs intrinsically suffer from significant optical loss. Even if plasmonic light trapping using gold (Au) and silver (Ag) NPs has attracted significant attentions regardless of solar cell type,[33-40] $\eta_{ext}$ is likely compromised through optical loss of these noble NPs in luminescence spectra.

As such, we used aluminum (Al) NPs as meta-atoms rather than conventional Au and Ag counterparts (see **Figure 1**b). This is because Al NPs exhibit significant optical loss mainly at ultraviolet (UV) wavelengths (less than 350 nm for 30 nm by 30 nm by 20 nm Al NPs) compared with the visible and near-infrared (IR) ranges.[41,42] Also, the dipolar resonance of Al NPs at UV wavelengths can still increase $n_{eff}$ in a quasi-static regime (including visible and near-IR) with negligible optical loss.[24-26]

For example, an effective optical medium in which 2D tetragonal arrays of 30 nm by 30 nm by 20 nm Al NPs with 10 nm gap are dispersed in 30 nm thick air shows an $n_{eff}$ of 6.8 at a dipolar resonance wavelength (340 nm) and that of 4.5 at the quasi-static limit (wavelengths of 400 ~ 1000 nm), both of which are much higher than the $n$ of the host medium (i.e., 1.0), as shown in **Figure 1**c. Details on numerical calculation of this effective parameter (i.e., the *s*-parameter retrieval method)[24,25] are summarized in Methods part. In particular, the figure of merit (FOM) was larger than 10 at the quasi-static limit (the imaginary part of $n$ was ranged from 0.15 to 0.21), as shown in **Figure S2**. As such, deep-subwavelength Al NPs can act as electric meta-atoms for enhancing the $n_{eff}$ of an optical effective medium in conjunction with minimal loss in luminescence spectra. Further increase in size of Al NPs from 30 nm by 30 nm by 20 nm could lead to the optical loss at the visible-near infrared regime, overlapped with luminescence spectra. In contrast, Au and Ag counterparts exhibit significant optical loss at the wavelength of 500 ~ 800 nm (see **Figure S3**), which could result in a decrease in $P_{esc}$ and the parasitic absorption of solar spectrum.

**Figure 1**d summarizes the $n_{eff}$ of an Al NP PMPEM. Herein, 30 nm by 30 nm by 20 nm Al NPs are assumed to be arrayed in a 2D tetragonal lattice with a different lateral gap spanning from 5 nm to 100 nm. Also, Al NPs are designed to be encapsulated in a 1 nm polymeric shell (an organic ligand with an $n$ of 1.45). The effective thickness of the PMPEM is found to be 30 nm, satisfying no explicit physical boundaries for retrieval of $n_{eff}$.[22] The $n$ of the perovskite (i.e., lead halide) host medium used in this study is shown in **Figure S4**. The smaller gap between Al NPs (higher vol% of Al NPs) gives rise to higher $n_{eff}$ at the wavelength of interest (beyond bandgap): at solar spectrum (particularly beyond bandgap), $n_{eff}$ can be broadly tuned from ~ 2.75 to ~ 6.00 with respect to vol% of Al NPs. Effective absorption ($k_{eff}$) of PMPEM with respect to vol% of Al NPs is shown in **Figure S5**. Even if we use the regularly arrayed Al NPs for clarity, the random dispersion of Al NPs can be also used to increase $n_{eff}$ (**Figure S6**).[26]

Fundamental efficiency limits and tradeoff problems: Using a detailed balance analysis, we systematically verified how a collective set of $a$, $J_{sc}$, $\eta_{ext}$, $V_{oc}$, fill factor (FF), and PCE is influenced by increasing the $n_{eff}$ of the active layer (**Figure 2**). The $n_{eff}$ of the active layer (i.e., PMPEM) used in this analysis, is summarized in **Figure 1**d. More information about the detailed balance analysis is included in Methods part. Thickness of the active layer varied from 30 nm to 1500 nm, as with previously reported detailed balance analysis on perovskite solar cells.[20] Also, SRH non-radiative recombination should be absent, when calculating fundamental limit of efficiency (i.e., $\tau$ is infinite). For the same reason, a perfect mirror with 100 % reflectivity ($R$) is used as a rear electrode. Other basic parameters of lead halide, including $i$, $C_{Aug}$, and $B_{int}$, are described in Methods part.

Several features are noteworthy, as follows. First, $a$ and $J_{sc}$ increase with thickness of the PMPEM (**Figures 2**a-b). This is because the increased thickness allows the active layer to absorb more light, according to equation (1). More importantly, it is obvious that higher $n_{eff}$ results in higher $a$ and the resulting $J_{sc}$, due to enhanced light trapping (from bluish to reddish lines). As such, plasmonic metamaterials appear to be a valid option to improve light trapping in perovskite solar cells.

Second, in contrast, $V_{oc}$ decreases for thicker solar cells (**Figures 2**c). Increasing $a$, particularly beyond the direct bandgap, by increasing the thickness of the active layer reduces the gap between $n$- and $p$-type quasi-Fermi levels (i.e., reducing the effective bandgap). $V_{oc,max}$ is tunable, because $V_{oc,max}$ relies on a quasi-Fermi level instead of an intrinsic bandgap. Also, $\eta_{ext}$ is reduced by increasing the thickness of the active layer (**Figures 2**d), because thicker active layers lead to a smaller $P_{esc}$:[21,43,44]

$$P_{esc} = \frac{\int_0^\infty \pi a \Phi_{BB} dE}{\int_0^\infty 4\pi n^2 \alpha d \Phi_{BB} dE} \tag{5}$$

where $\Phi_{BB}$ indicates blackbody radiation spectral density. These two effects synergistically decrease $V_{oc}$ when the thickness of the active layer is increased.

Furthermore, an increase in $n_{eff}$ reduces $\eta_{ext}$ and thus $V_{oc}$, because an enhanced $n_{eff}$ narrows the escape cone and thereby further decreases $P_{esc}$ (see equation 5). $\Phi_{BB}$ is not influenced by $n_{eff}$, as follows:

$$\Phi_{BB}(E) = \frac{2n_{sub}^2 E^2}{h^3 c^2} \exp\left(-\frac{E}{kT}\right) \tag{6}$$

where $h$, $c$, and $n_{sub}$ are Plank's constant, the speed of light, and $n$ of the substrate, respectively. $P_{esc}$ herein is solely affected by $n_{eff}$.

Note that the $n^2$ term in the denominator of equation (5) accounts for photon-recycling;[43,44] photon-recycling cannot affect $J_{sc}$. Thus, equation (5), recently derived by U. Rau et al.,[43,44] indicates that $\eta_{ext}$ is a gauge not only of loss mechanisms (i.e., $A_{SRH}$, $C_{Aug}$, and $P_{par}$), but also of the degree of photon-recycling. Given this claim, we conclude that greater photon-recycling results in poorer $P_{esc}$, which in turn decreases $\eta_{ext}$ and $V_{oc}$. As recently described by Yablonovitch and his colleagues,[21] a higher $n_{eff}$ and the resulting more crowd photon gas can boost the luminescence emission rate rather than $\eta_{ext}$. Also, because absorption should be balanced by emission,[27,28] less-probable $\eta_{ext}$ resulting from higher $n_{eff}$ restricts an accessible amount of both light absorption and photo-generated carriers. Overall, as the density of photo-generated carriers defines voltage, boosting photon-recycling by increasing $n_{eff}$ results in a reduction in $V_{oc}$.

In addition, the rates of $\eta_{ext}$ and $V_{oc}$ decrease, caused by enhancing $n_{eff}$, are accelerated with increasing thickness of the active layer (see **Figures 2**d). This means that the effect of photon-recycling is also enhanced with increasing thickness of the active layer. In addition to photon-recycling, the effective bandgap of perovskite is narrowed by increasing the thickness-enabled enhancement of light trapping, as mentioned above. These two effects cooperatively reduce $V_{oc}$. As a result, $V_{oc}$ and $J_{sc}$ trade off with each other, once both light trapping and photon-recycling are simultaneously boosted by increasing $n_{eff}$.

These results provide an important insight into potential new directions for plasmonic metamaterial solar cell design. Previous reports on embedding plasmonic NPs into active layers of solar cells focused mainly on near-field enhancement by localized surface plasmon resonance (LSPR) and its use in increasing overall $a$ (i.e., antenna effect).[32-40] This LSPR effect was also used for improvement of light absorption in perovskite solar cells.[39,45-47] However, LSPR increases not only near-field absorption, but also $n_{eff}$.[24-26] Therefore, thermodynamic change caused by an increasing $n_{eff}$ needs to be fully reflected to accurately define the fundamental limit of efficiency, as detailed above.

Meanwhile, Snaith et al. suggested that inclusion of plasmonic NPs (i.e., Ag NPs) in perovskite could result in enhanced radiative decay of excitons and therefore an increase in photocurrent.[48,49] This effect could further improve luminescence emission rates and the amount of photo-generated carriers. We focus on the thermodynamic variations of solar cells caused by increasing the $n_{eff}$ of the perovskite active layer, but the possibility that such electronic effects arise from plasmonic NPs should be studied further and included in future theoretical analyses on the fundamental limit of plasmonic perovskite solar cells.

**Figures 2**e-f show FF and PCE variations according to $n_{eff}$ of PMPEM. With and without Al NPs, PCE is enhanced with increasing thickness of the PMPEM. More importantly, despite the $V_{oc}$ drop, PCE can still be enhanced by increasing $n_{eff}$ (FF slightly decreases). This enhancement of PCE by increasing $n_{eff}$ is visible regardless of PMPEM thickness. This result implies the resultant $J_{sc}$ boost can compensate for the $V_{oc}$ decrease.

**Practical limitations caused by non-radiative recombination**: Next, we explore how the material quality of lead halide affects the efficiency of plasmonic metamaterial perovskite solar cells. Defects in lead halide give rise to trap-assisted non-radiative recombination, significantly reducing the non-radiative lifetime ($\tau$) in a realistic condition.[19-21] According to previous reports, the experimentally accessible maximum $\tau$ of perovskite is approximately 1 μs.[12,50,51] To define the practical limit of efficiency, we take a $\tau$ of 1 μs in the detailed balance analyses, while the $R$ of a rear mirror is still 100 % (perfect mirror).

**Figure 3** and **Figure S6** present the corresponding results of $a$, $J_{sc}$, $\eta_{ext}$, $V_{oc}$, fill factor (FF), and PCE as functions of $n_{eff}$ and perovskite thickness. The values of $a$ and $J_{sc}$ remain intact even with a distinct non-radiative recombination (see **Figure S7**). This is because a non-radiative recombination affects only $\eta_{ext}$ and thus $V_{oc}$, while $J_{sc}$ is a function of $a$ (see equations of 1-4). The cases with and without Al NPs show the same variations of $a$ and $J_{sc}$ with respect to $n_{eff}$ and PMPEM thickness (**Figure S7**).

Critically, as shown in **Figures 3**a-b, the reduced $\tau$ (from infinite to 1 μs) causes both $\eta_{ext}$ and $V_{oc}$ to decrease dramatically (e.g., from 1.27 V to 1.17 V for 1000 nm thick bare perovskite without plasmonic metamaterials). More importantly, the degrees of falls in $\eta_{ext}$ and $V_{oc}$, driven by increasing the thickness of the active layer, becomes increased after reducing $\tau$ from infinite to 1 μs (compare **Figure 3**a with **Figure 2**d). Non-radiative recombination is a bulk effect of materials; it plays a bigger role in defining $\eta_{ext}$ and $V_{oc}$ with increasing the active layer thickness. Consequently, a significantly lower $V_{oc}$, particularly for a relatively thicker active layer, cannot be overcome, even by enhanced light trapping; PCE of bare perovskite solar cells, handicapped with a non-radiative recombination, peaks at a relatively thin active layer (26.9 % for 200 nm thickness) and gradually decreases beyond that thickness (**Figures 3**c). Corresponding FF data are included in **Figure S8**. This PCE trend with respect to thickness of the active layer contrasts starkly with results achieved without a non-radiative penalty (compare **Figure 3**c with **Figure 2**f).

As with an ideal perovskite, increasing the $n_{eff}$ of realistic perovskite significantly reduces $V_{oc}$ due to the narrowed effective bandgap and a decrease in $\eta_{ext}$. More critically, non-radiative recombination further facilitates the degree of $V_{oc}$ drop at each thickness (**Figure 3**b). In a realistic situation, an improved $J_{sc}$ cannot compensate for the $V_{oc}$ drop via increasing $n_{eff}$, especially for a relatively thick perovskite (i.e., thicker than 600 nm). For the active layers thicker than 600 nm, the PCE of plasmonic metamaterial perovskite solar cells becomes smaller than that of a bare perovskite counterpart. However, it is clear that increasing $n_{eff}$ can still improve the practical limit of PCE for relatively thin perovskite solar cells (< 600 nm); a PCE with a 5 nm gap-based PMPEM can peak (~ 27.8 %) at the thickness of 60 nm. This is because a $V_{oc}$ drop resulting from increasing $n_{eff}$ is relatively mitigated as the active layer is thinned. Indeed, the enhancement factor of PCE via increasing $n_{eff}$ (difference of PCE between metamaterial (PCE$_{MM}$) and pure (PCE$_{Bare}$) perovskite solar cells) becomes more significant as the thickness of perovskite is reduced (**Figures 3**d). As a result, we can achieve an increase in maximum PCE (from 26.9 to 27.8) even with a thinner perovskite solar cell (from 200 nm to 60 nm) by using plasmonic metamaterials in realistic situations.

**Practical limitations due to use of a realistic mirror**: In a realistic situation, the $R$ of a rear mirror is less than 100 %. A flat Au back electrode is located at the bottom of perovskite solar cell,[6-18] serving as a rear reflector. To optimize the carrier extraction, Au was chosen as a rear mirror element. However, parasitic absorption via surface plasmon resonance and interband transition inevitably arises from the use of flat Au, as shown in **Figure 4**a; the reflection spectra of a flat Au rear mirror used in this study is from experimental data.[52] Beyond an interband transition regime (i.e., wavelengths > 550 nm), the $R$ of a flat Au is approximately 97 ~ 99 %, which decreases significantly below 50 % at the interband transition regime (i.e., wavelengths < 500 nm). Thus, $a$ approximated by the 4n$^2$ limit should be modified to reflect the unperfected $R$ of a mirror:[21]

$$a(E) = \frac{4n^2 \alpha d}{4n^2 \alpha d + 1 + n^2(1-R)} \tag{7}$$

Furthermore, $P_{esc}$ can be handicapped as follows:[21]

$$P_{par} = P_{esc} \times n^2(1-R) \tag{8}$$

Because both $a$ and $P_{esc}$ are damaged by a factor of $n^2$ under realistic mirror conditions, an increase in $n_{eff}$ dilutes the effect of light trapping, while simultaneously reducing $\eta_{ext}$ and $V_{oc}$ further. **Figure 4**b summarizes how an unperfected mirror affects $a$ by comparing it with a perfected mirror. For both cases of real and ideal mirrors, τ is set to 1 μs, while for clarity, PMPEM with a gap of 5 nm is only included as a representative example of plasmonic metamaterial solar cells along with bare perovskite solar cells. As expected, the unperfected $R$ of a mirror leads to a reduction in $a$. Also, as shown in **Figure 4**c, the degree of light trapping through increasing $n_{eff}$ (i.e., the degree of the enhanced $a$, defined by $a_{MM} − a_{Bare}$) is diluted by $P_{par}$ of a real mirror and increases with reducing active layer thickness. The corresponding $J_{sc}$ is summarized in **Figure S9**.

Additionally, an unperfected $R$ of a mirror reduces $\eta_{ext}$ (see **Figure 4**d). More importantly, increasing the $n_{eff}$ further decreases $\eta_{ext}$ due to enhanced light trapping and photon-recycling; the degree of such fall in $\eta_{ext}$ caused by an increasing $n_{eff}$ becomes more significant for a thinner active layer; the corresponding $V_{oc}$ is summarized in **Figure S10**. This in turn further narrows the range of active layer thickness, in which an increase in $n_{eff}$ can enhance PCE. As shown in **Figure 5**a, for an active layer thinner than 330 nm, plasmonic metamaterial perovskite solar cells can outperform bare perovskite counterparts; corresponding FF data are summarized in **Figure S11**. Also, the degree of PCE enhancement by increasing $n_{eff}$ ($PCE_{MM}$-$PCE_{Bare}$) increases as the active layer is thinned (**Figure 5**b). This suggests that plasmonic metamaterials would be effective only for ultrathin perovskite solar cells (thinner than 100 nm); this restriction originates with the tradeoff between $V_{oc}$ and $J_{sc}$, in which light trapping and photon-recycling are simultaneously boosted by increasing $n_{eff}$. The random dispersion of Al NPs in perovskite solar cells also resulted in a similar performance enhancement for ultrathin active layer, as summarized in **Figure S12**.

Conclusions: We systematically verified how increasing the $n_{eff}$ of the active layer affects the performances of perovskite solar cells. Using the concept of plasmonic metamaterials, a theoretical strategy for increasing the $n_{eff}$ of perovskite is suggested. From a materialization perspective, our design could be readily accessible via random dispersing or entropic packing of Al NPs within perovskite media.[26] A detailed balance analysis was then carried out to determine the effect of increased $n_{eff}$ on solar cell efficiency. An off-resonant, broadband increase in $n_{eff}$ of PMPEM was found to enhance photon management of perovskite solar cells with minimal parasitic absorption. In that context, both $a$ and $J_{sc}$ can benefit. However, the enhanced photon-recycling via increasing $n_{eff}$ negatively affects $\eta_{ext}$ and the resulting $V_{oc}$. Despite the penalty of a $V_{oc}$ drop, the fundamental limits of PCE in perovskite can in principle be enhanced by increasing $n_{eff}$ regardless of the thickness of the active layer. In a realistic situation defined with a non-radiative recombination and an unperfected rear mirror, an increase in $n_{eff}$ holds promise, particularly for an ultrathin perovskite solar cell.

Methods

*Detailed balance analysis*: In detailed balance analysis, the total current density ($J_{total}$) is calculated by generation and recombination rates expressed with current density. These current density components are combined as followings:

$$J_{total}(V) = J_{sc} - J_{0,rad}(V) - J_{Auger}(V) - J_{SRH}(V) \tag{9}$$

The three loss mechanisms of perovskite solar cell are considered: radiative recombination rate ($J_{0,rad}$), auger recombination rate ($J_{Auger}$), Shockley-Read-Hall recombination rate ($J_{SRH}$). Each of these terms are defined by:

$$J_{sc} = q \int_0^\infty a(E)\Phi_S(E)dE \tag{10}$$

$$J_{0,rad}(V) = q\pi e^{\frac{qV}{kT}} \int_0^\infty a(E)\Phi_{BB}(E)dE \tag{11}$$

$$J_{Auger}(V) = qe^{\frac{3qV}{2kT}} C_{Auger} i^3 d \tag{12}$$

$$J_{SRH}(V) = qe^{\frac{qV}{2kT}} A_{SRH} i d \qquad (13)$$

where $a(E)$, $\Phi_S$, $\Phi_{BB}$, $C_{Auger}$, $i$, d, and $A_{SRH}$ are the absorptivity, solar spectrum, black body radiation of the solar cell, auger recombination coefficient, intrinsic carrier concentration, thickness of solar cell, and Shockley-Read-Hall recombination coefficient respectively. The value of $B_{int}$, $C_{Auger}$ are $1.34 \times 10^{-10}$ (cm$^3$/s), $1.1 \times 10^{-28}$ (cm$^6$s$^2$) respectively.[23] The intrinsic carrier density($i$) is estimated from van Roosbroeck-Shockley Relationship:[53]

$$B_{int} i^2 = \int_0^\infty 4\pi n^2 \alpha \Phi_{BB} dE \qquad (14)$$

*Numerical calculations of effective parameters of a PMPEM*: Scattering parameters including transmission, reflection, and relevant phase changes are numerically calculated by finite-difference, time-domain (FDTD), supported by a commercial package (CST Microwave Studio 2014). The dielectric constants of Al, Au, and Ag were derived from a Drude-critical model. A periodic boundary condition was used; perfect matched layers (PML) were placed at the top and bottom of the simulation model to avoid reflection from the boundaries. We then obtained all effective parameters using an *s*-parameter extraction method based on homogenization theory.[24-26] Both the thickness of metamaterials (30 nm) and size of meta-atoms (plasmonic NPs) were deep-subwavelength scale (~ $0.37\lambda/n_{eff}$), so as to justify the use of homogenization theory.

Acknowledgement

This work was supported by Samsung Research Funding Center for Samsung Electronics under Project Number SRFC-MA1402-09.

Supporting Information

Details about (1) index-engineering for light trapping, (2) FOM of Al NPs-based optical effective medium (host medium of air), (3) optical effective medium consisting of gold (Au) and silver (Ag) NPs (host medium of air), (4) $n$ of lead halide host medium used in this work, (5) $k_{eff}$ of Al plasmonic metamaterial perovskite effective medium, (6) $n$ and solar cell performance of the randomly dispersed Al NPs-perovskite solar cells, (7) $a$, $J_{sc}$, and FF of the Al plasmonic metamaterial perovskite solar cells, handicapped with a non-radiative recombination, and (8) $J_{sc}$, $V_{oc}$, and FF of the Al plasmonic metamaterial perovskite solar cells, handicapped with a non-radiative recombination and a realistic rear mirror are available in Supporting Information. This material is available free of charge *via* the internet at http://pub.acs.org.

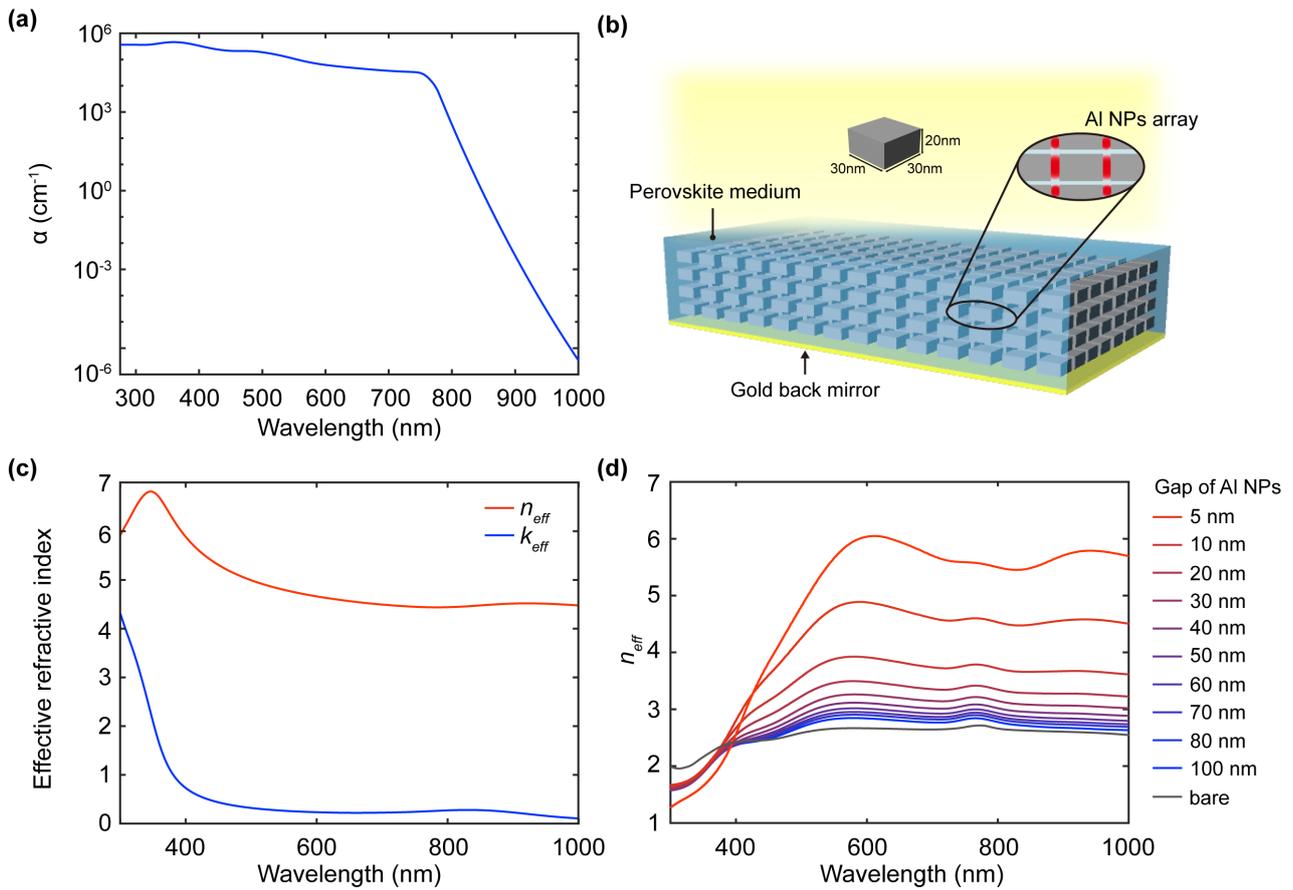

**Figure 1.** (a) Absorption coefficient of perovskite (e.g., lead halide) used in this study. (b) Schematic for plasmonic metamaterial perovskite solar cells. (c) Effective refractive index ($n_{\text{eff}}$ and $k_{\text{eff}}$) of plasmonic metamaterial effective medium, where air is the host materials, and 30 nm by 30 nm by 20 nm aluminum (Al) nanoparticles (NPs) are assumed to be arrayed with a 10 nm gap. (d) $n_{\text{eff}}$ of a plasmonic metamaterial perovskite effective medium (PMPEM), where 30 nm by 30 nm by 20 nm Al NPs are assumed to be arrayed with a different gap spanning from 5 nm to 100 nm. Complex dielectric constants of perovskite used in this calculation are presented in **Figure S2**, Supporting Information.

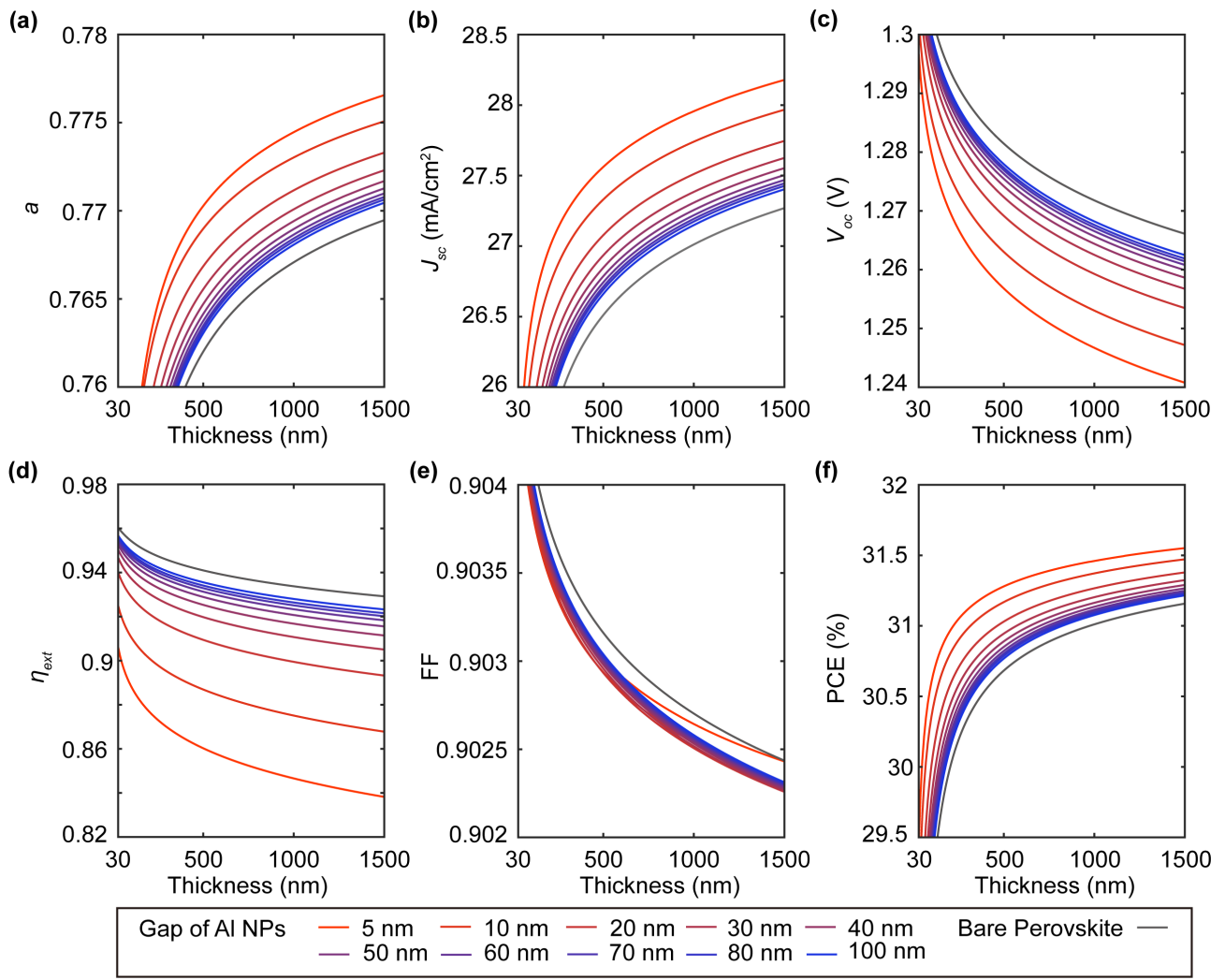

**Figure 2.** Fundamental limit of perovskite solar cells with and without Al NPs. From bluish to reddish colors, $n_{eff}$ of PMPEM increases according to vol% of Al NPs, as noted in **Figure 1**d. Thickness of the active layer varies from 30 nm to 1500 nm. (a) Absorptivity ($a$). (b) Short-circuit current ($J_{sc}$). (c) Open-circuit voltage ($V_{oc}$). (d) External luminescence efficiency ($\eta_{ext}$). (e) Fill factor (FF). (f) Power conversion efficiency (PCE, %).

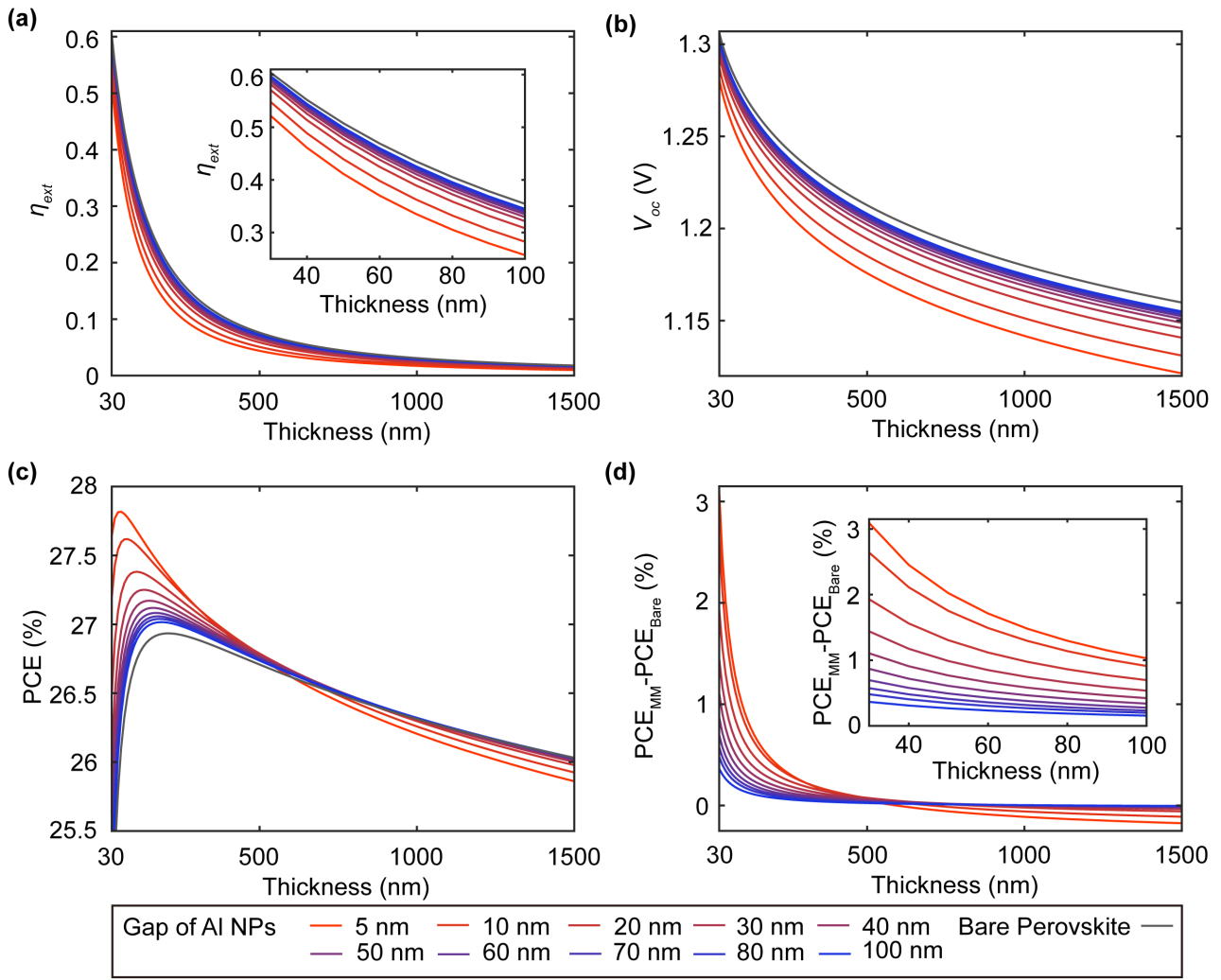

**Figure 3.** Practical limit part 1: PMPEM solar cells with a non-radiative recombination (non-radiative life time (τ) is set to 1 μs). From bluish to reddish colors, $n_{eff}$ of PMPEM increases according to vol% of Al NPs, as noted in **Figure 1**d. (a) $\eta_{ext}$. (b) $V_{oc}$. (c) PCE. (d) PCE enhancement by increasing $n_{eff}$ (i.e., $PCE_{MM}$ (PMPEM) – $PCE_{Bare}$ (bare perovskite)).

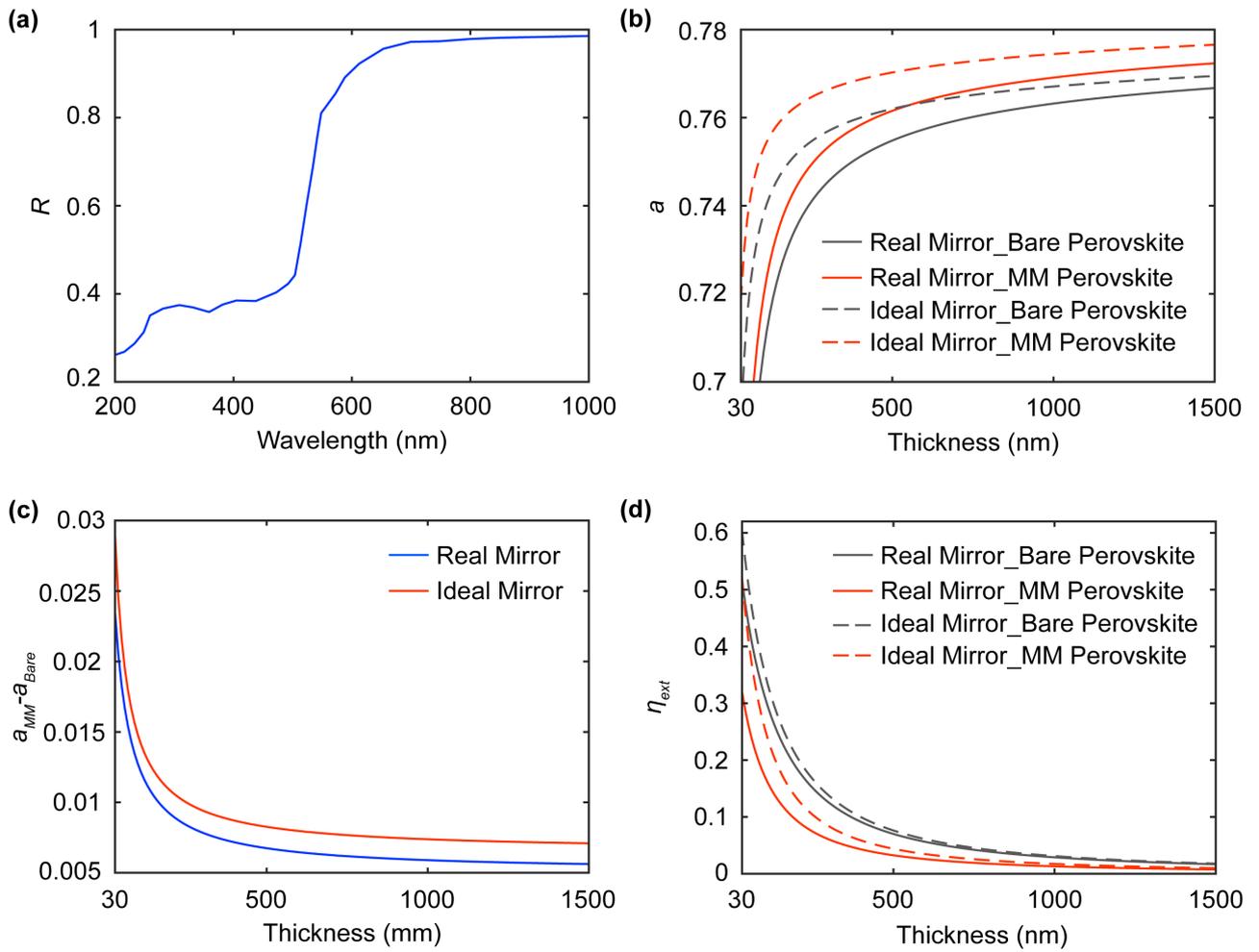

**Figure 4.** Practical limit part 2: PMPEM solar cells with a non-radiative recombination (τ is set to 1 μs) and a realistic rear mirror (a flat Au mirror). For clarity, the results of 5 nm gap PMPEM (legend of MM Perovskite) and bare perovskite solar cells are included. (a) Reflectivity ($R$) of a realistic rear mirror used in this work. (b) Absorptivity ($a$) comparison between ideal (bold line) and realistic (dotted line) rear mirrors. (c) Enhancment of absorptivity ($a_{MM}$-$a_{Bare}$) by increasing $n_{eff}$ for an ideal (red line) and realistic mirror (blue line). (d) $\eta_{ext}$ comparison between ideal (bold line) and realistic (dotted line) rear mirrors. The thickness of the active layer varies from 30 nm to 1500 nm (b-d).

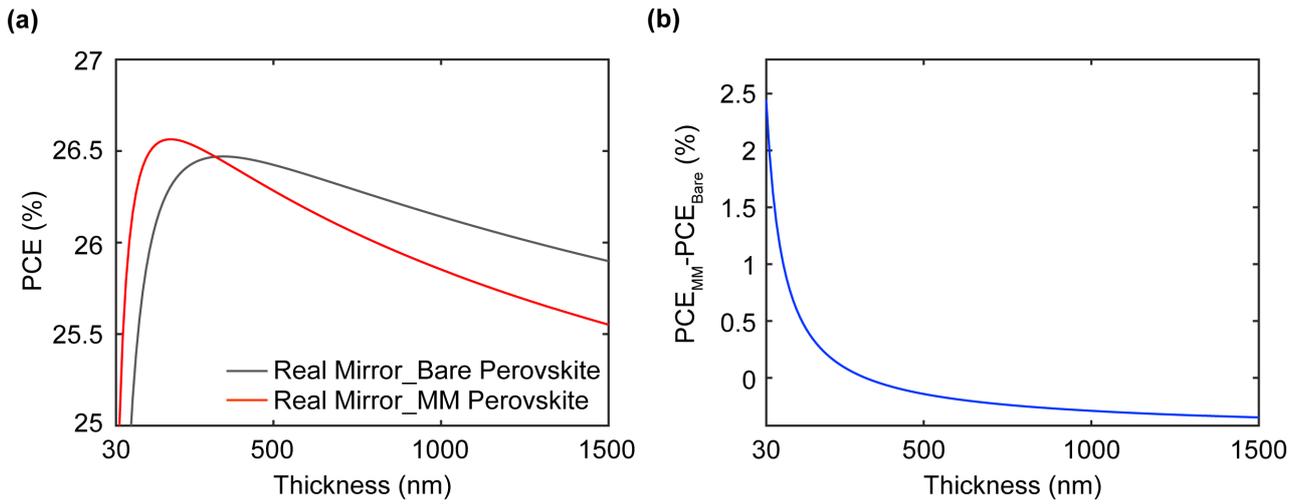

**Figure 5.** Practical limit part 2: PCE analysis on realistic PMPEM solar cells with non-radiative recombination ($\tau$ is set to be 1 μs) and a realistic rear mirror (a flat Au mirror). (a) PCE comparison between perovskite solar cells with (red line) and without (gray line) PMPEM. (b) PCE enhancement by increasing $n_{eff}$ (i.e., $PCE_{MM}$ (PMPEM) – $PCE_{Bare}$ (bare perovskite). The thickness of the active layer varies from 30 nm to 1500 nm (a-b).

*Supporting Information for*

# Plasmonic Metamaterial Perovskite Solar Cells: Fundamental Tradeoffs, Limitations, and Opportunities

*Kwangjin Kim*[1,2] and *Seungwoo Lee*[2*]

[1]SKKU Advanced Institute of Nanotechnology, Sungkyunkwan University (SKKU), Suwon 16419, Republic of Korea
[2]KU-KIST Graduate School of Converging Science and Technology, Korea University, Seoul 02841, Republic of Korea

*Email: seungwoo@korea.ac.kr

Contents:

1. Index-engineering for light trapping

2. Figure-of-merit (FOM) of Aluminum (Al) nanoparticles (NPs)-based optical effective medium (host medium of air)

3. Optical effective medium consisting of gold (Au) and silver (Ag) NPs (host medium of air)

4. Refractive index of lead halide host medium, used in this work

5. Im($n$) of Al plasmonic metamaterial perovskite effective medium

6. Enhancement of $n_{eff}$ by the random dispersion of Al NPs

7. $a$ and $J_{sc}$ of the Al plasmonic metamaterial perovskite solar cells, handicapped with a non-radiative recombination

8. Fill factor (FF) of the Al plasmonic metamaterial perovskite solar cells, handicapped with a non-radiative recombination

9. $J_{sc}$ of the Al plasmonic metamaterial perovskite solar cells, handicapped with a non-radiative recombination and a realistic rear mirror

10. Open circuit voltage ($V_{oc}$) of the perovskite solar cells, handicapped with a non-radiative recombination

11. FF of the perovskite solar cells, handicapped with a non-radiative recombination and a realistic rear mirror

12. $J_{sc}$, $V_{oc}$, and PCE of the randomly dispersed Al NSs-perovskite solar cell under realistic condition

## 1. Index-engineering for light trapping

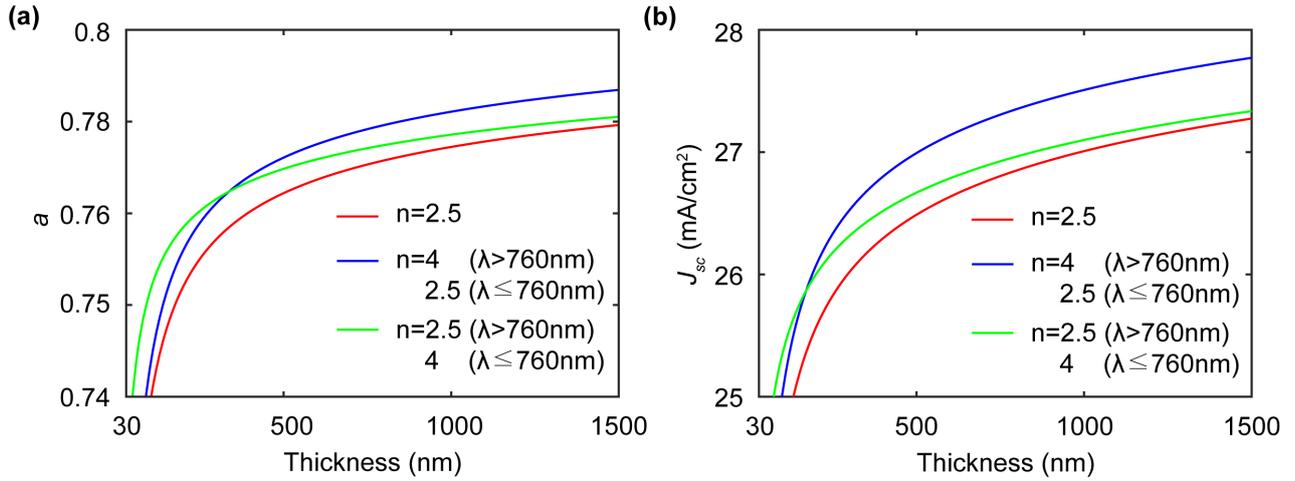

**Figure S1.** Effect of refractive index engineering on light absorptivity ($a$) (a) and short circuit current ($J_{sc}$). Herein, the active layer of solar cell is assumed to have a constant refractive index of 2.5 without dispersity. Also, absorption coefficient ($\alpha$) of this active layer is taken from Figure 1(a) of main manuscript.

2. Figure-of-merit (FOM) of Aluminum (Al) nanoparticles (NPs)-based optical effective medium (host medium of air)

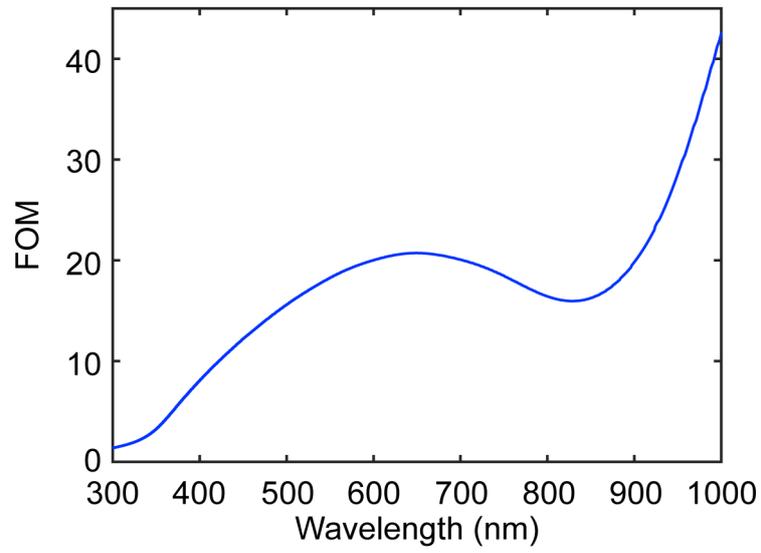

**Figure S2.** FOM ($n_{eff}/k_{eff}$) of Al NPs-based optical effective medium: host medium is air with *n* of 1.

## 3. Optical effective medium consisting of gold (Au) and silver (Ag) NPs (host medium of air)

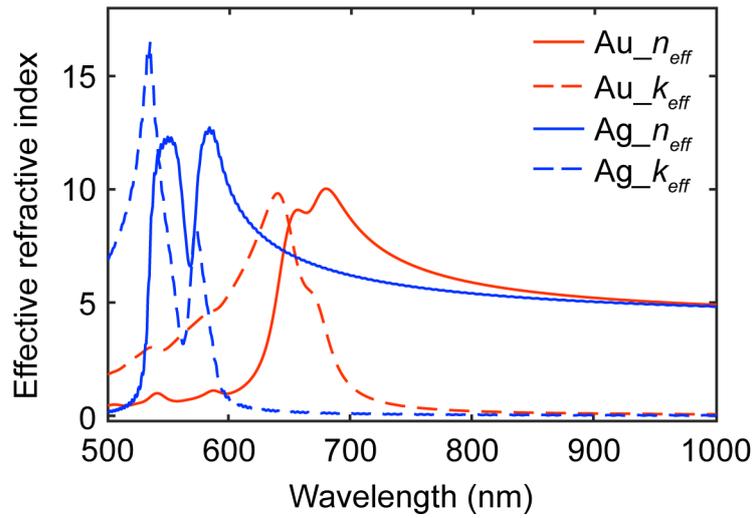

**Figure S3.** Effective refractive index (real part: $n_{eff}$, imaginary part: $k_{eff}$) of 30 nm by 30 nm by 20 nm Au and Ag NPs-dispersed air medium, which is obtained by *s*-parameter retrieval method.[23,24] Herein, Au and Ag NPs are regularly arrayed in 2D tetragonal lattice with 10 nm gap. Effective thickness of air host medium is 30 nm, which justify no explicit physical boundary for retrieval of $n_{eff}$ and $k_{eff}$.

4. Refractive index of lead halide host medium, used in this work

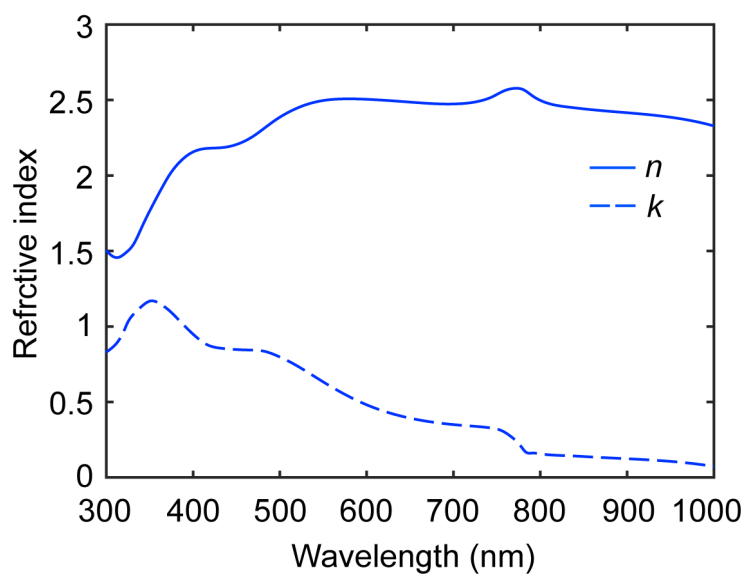

**Figure S4.** Real (*n*) and imaginary (*k*) refractive index of perovskite used in this study. This data is taken from (J. M. Ball et al., *Energy Environ. Sci.* **2015**, *8*, 602−609).

## 5. Absorption of Al plasmonic metamaterial perovskite effective medium

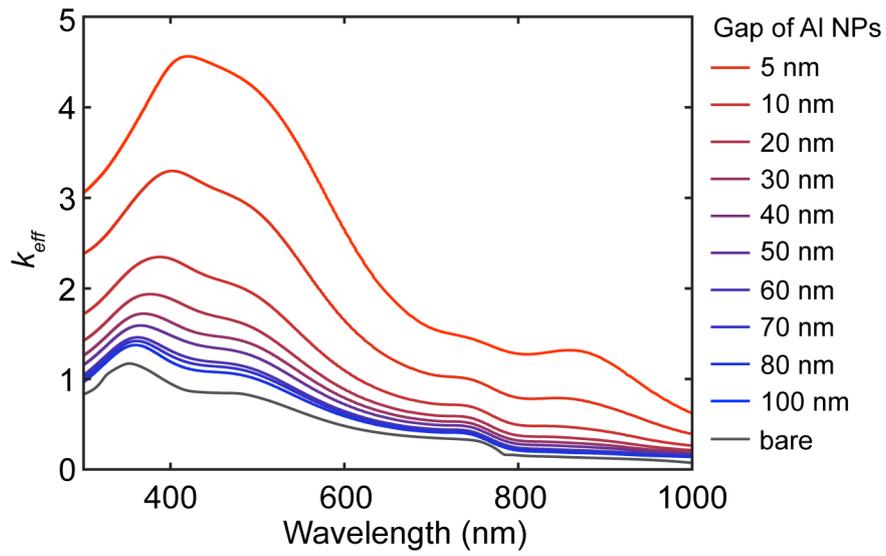

**Figure S5.** Absorption of Al plasmonic metamaterial perovskite effective medium with a different gap between Al NPs: herein, perovskite is a host medium; corresponding $n_{eff}$ is included in Figure 1(d) of main manuscript.

## 6. Enhancement of $n_{eff}$ by the random dispersion of Al NPs

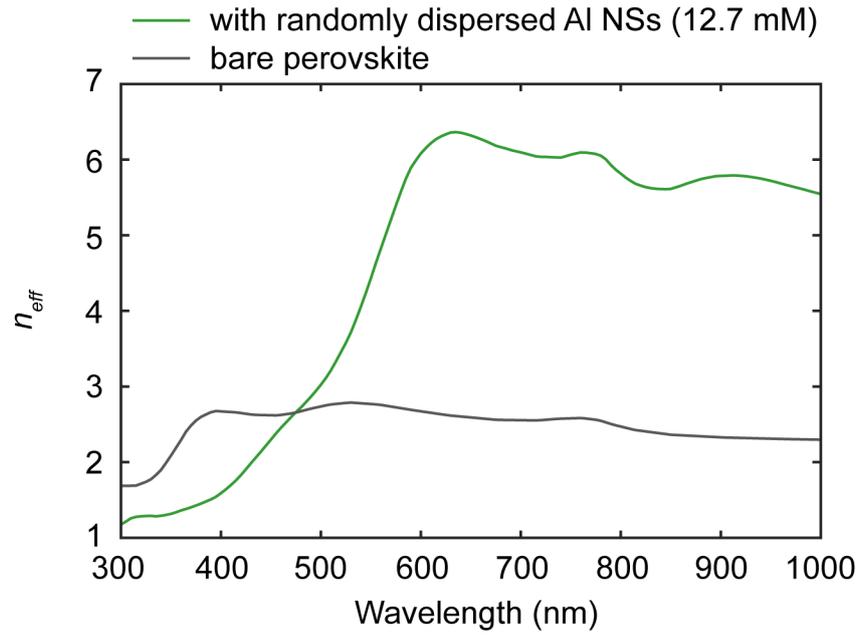

**Figure S6.** The enhancement of $n_{eff}$ by a random dispersion of Al NPs in perovskite medium. Herein, Al NPs were assumed to be nanospheres (NSs) with 5 nm size; the concentration of Al NSs was 12.7 mM concentration.

7. *a* and *J*<sub>sc</sub> of the Al plasmonic metamaterial perovskite solar cells, handicapped with a non-radiative recombination

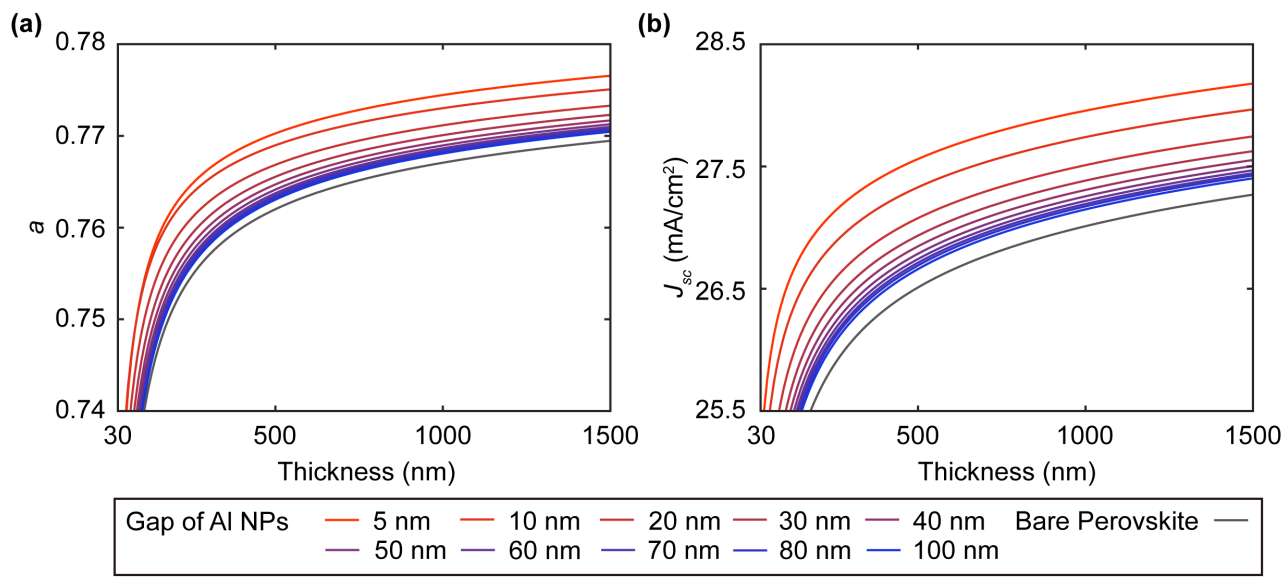

**Figure S7.** (a) *a* and (b) *J*$_{sc}$ of Al plasmonic metamaterial perovskite solar cells, handicapped with a non-radiative recombination. Herein, non-radiative lifetime is set by 1 µs, while reflectivity of a rear mirror (*R*) is designed to be 100 % (i.e., perfect mirror).

8. Fill factor (FF) of the Al plasmonic metamaterial perovskite solar cells, handicapped with a non-radiative recombination

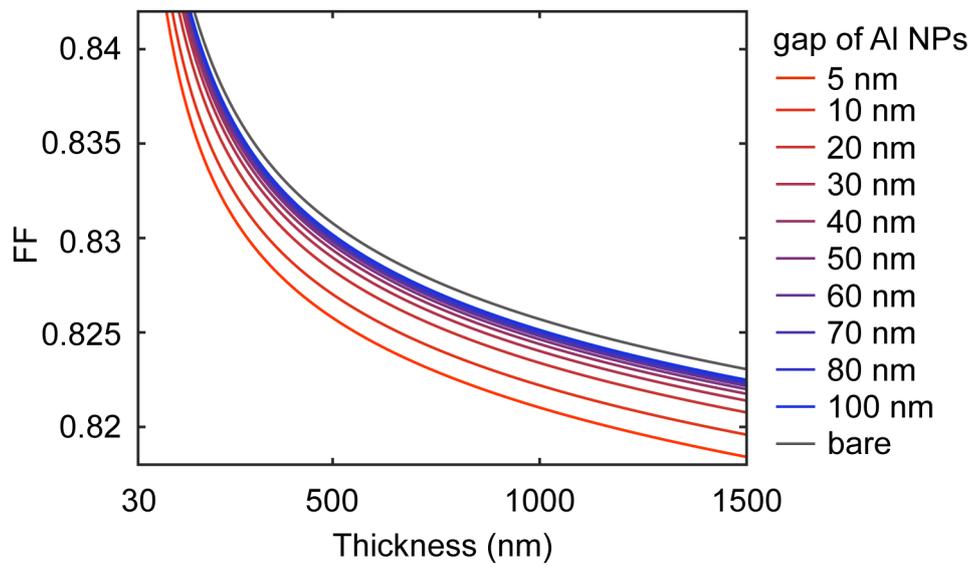

**Figure S8.** FF of Al plasmonic metamaterial perovskite solar cells, handicapped with a non-radiative recombination. Herein, non-radiative lifetime is set by 1 μs, while reflectivity of a rear mirror ($R$) is designed to be 100 % (i.e., perfect mirror).

9. $J_{sc}$ of the Al plasmonic metamaterial perovskite solar cells, handicapped with a non-radiative recombination and a realistic rear mirror

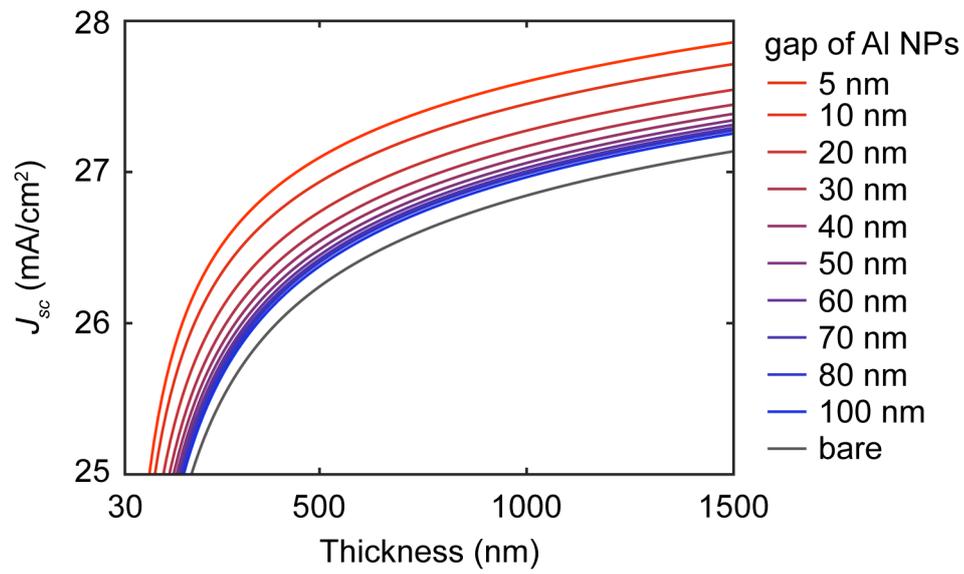

**Figure S9.** $J_{sc}$ of Al plasmonic metamaterial perovskite solar cells, handicapped with a non-radiative recombination (non-radiative lifetime is set by 1 µs) and unperfected mirror.

10. Open circuit voltage ($V_{oc}$) of the perovskite solar cells, handicapped with a non-radiative recombination

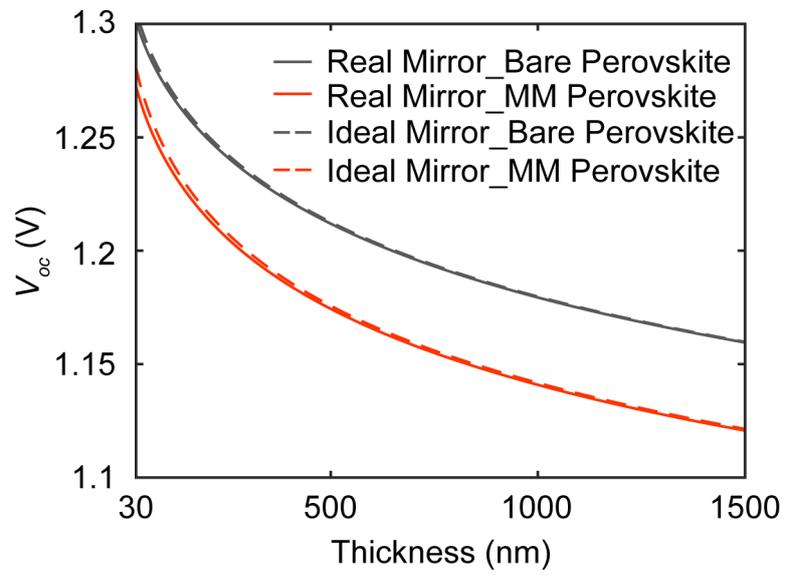

**Figure S10.** $V_{oc}$ of bare and Al plasmonic metamaterial perovskite solar cells, handicapped with a non-radiative recombination (non-radiative lifetime is set by 1 μs). Ideal (with $R$ of 100 %) and unperfected real (with $R$, presented in Figure 4a of main manuscript) mirrors are used.

11. FF of the perovskite solar cells, handicapped with a non-radiative recombination and a realistic rear mirror

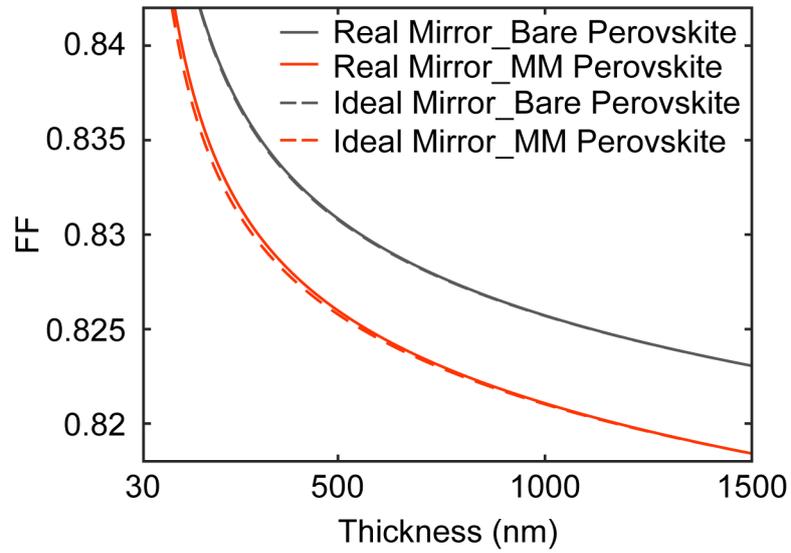

**Figure S11.** FF of Al plasmonic metamaterial perovskite solar cells, handicapped with a non-radiative recombination (non-radiative lifetime is set by 1 μs). Ideal (with *R* of 100 %) and unperfected real (with *R*, presented in Figure 4a of main manuscript) mirrors are used.

12. $J_{sc}$, $V_{oc}$, and PCE of the randomly dispersed Al NSs-perovskite solar cell under realistic condition

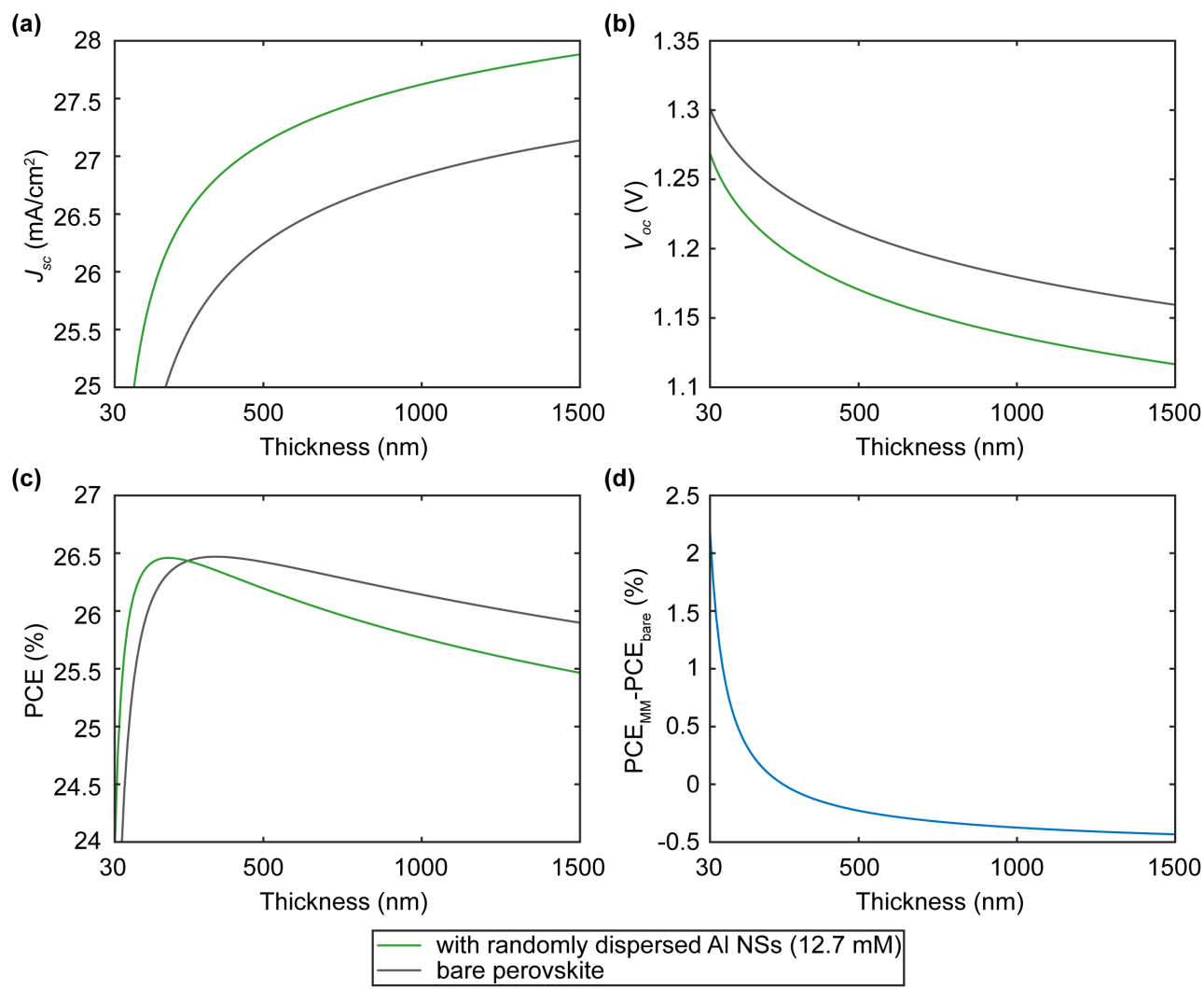

**Figure S12.** (a) $J_{sc}$, (b) $V_{oc}$, and (c) PCE comparison between perovskite solar cells with (light green line) and without (gray line) Al NSs (5 nm sized and 12.7 mM concentrated random dispersion). The $n_{eff}$ of Al NS-randomly dispersed perovskite is described in Figure S6. The realistic limits such as a real Au mirror and $10^{-6}$ of SRH lifetime were reflected in this calculation. (d) PCE enhancement by increasing $n_{eff}$ (i.e., $PCE_{MM}$ (with random dispersion of Al NSs) – $PCE_{Bare}$ (bare perovskite). The thickness of the active layer varies from 30 nm to 1500 nm (a-d).